\documentclass[onecolumn,aps,superscriptaddress,amsmath,secnumarabic,pdflatex,
amssymb,prl,preprint]{revtex4}
\usepackage{bm}
\usepackage{amsmath,amsfonts,amssymb}
\usepackage{epstopdf}
\usepackage{xcolor}
\usepackage{color}
\usepackage{wasysym}

\usepackage{graphicx}

\begin{document}
\title{Brilliant GeV Gamma-ray flash from Inverse Compton Scattering in QED Regime}

\author{Z. Gong}
\affiliation{State Key Laboratory of Nuclear Physics and Technology, and Key Laboratory of HEDP of the Ministry of Education, CAPT, Peking University, Beijing 100871, China}
\author{R. H. Hu}
\affiliation{State Key Laboratory of Nuclear Physics and Technology, and Key Laboratory of HEDP of the Ministry of Education, CAPT, Peking University, Beijing 100871, China}
\author{H. Y. Lu}
\affiliation{State Key Laboratory of Nuclear Physics and Technology, and Key Laboratory of HEDP of the Ministry of Education, CAPT, Peking University, Beijing 100871, China}
\author{J. Q. Yu}
\affiliation{State Key Laboratory of Nuclear Physics and Technology, and Key Laboratory of HEDP of the Ministry of Education, CAPT, Peking University, Beijing 100871, China}
\author{D. H. Wang}
\affiliation{State Key Laboratory of Nuclear Physics and Technology, and Key Laboratory of HEDP of the Ministry of Education, CAPT, Peking University, Beijing 100871, China}
\affiliation{State Key Laboratory of Laser Interaction with Matter, Northwest Institute of Nuclear Technology, Xi'an, 710024, China}
\author{E. G. Fu}
\affiliation{State Key Laboratory of Nuclear Physics and Technology, and Key Laboratory of HEDP of the Ministry of Education, CAPT, Peking University, Beijing 100871, China}
\author{C. E. Chen}
\affiliation{State Key Laboratory of Nuclear Physics and Technology, and Key Laboratory of HEDP of the Ministry of Education, CAPT, Peking University, Beijing 100871, China}
\author{X. T. He}
\affiliation{State Key Laboratory of Nuclear Physics and Technology, and Key Laboratory of HEDP of the Ministry of Education, CAPT, Peking University, Beijing 100871, China}
\author{X. Q. Yan\footnotetext{$\dagger$ x.yan@pku.edu.cn}}
 \email[]{x.yan@pku.edu.cn}
 \affiliation{State Key Laboratory of Nuclear Physics and Technology, and Key Laboratory of HEDP of the Ministry of Education, CAPT, Peking University, Beijing 100871, China}
 \affiliation{Collaborative Innovation Center of Extreme Optics, Shanxi University, Taiyuan, Shanxi 030006, China}
 \affiliation{Shenzhen Research Institute of Peking University, Shenzhen 518055, China}

\date{\today}

\begin{abstract}
An all-optical scheme is proposed for studying a laser-plasma based incoherent photon emission from inverse Compton scattering in quantum electrodynamic (QED) regime. A theoretical model is presented to explain the coupling effect among radiation reaction trapping, self-generated magnetic field and spiral attractor in phase space, which guarantees the energy and angular momentum (AM) transformation from electromagnetic fields to particles. 
Taking advantage of a prospective $\sim$10$^{23}$W/cm$^{2}$ laser facility, 3D Particle-in-cell (PIC) simulations manifest the present gamma-ray flash with an unprecedented power of multi-petawatt (PW) and brightness of 1.7$\times$10$^{23}$photons/s/mm$^2$/mrad$^2$/0.1$\%$bandwidth (at 1GeV). These results bode well for new research direction in particle physics and laboratory astrophysics while exploring laser plasma interaction.
\end{abstract}

\maketitle
\subsection{1. Introduction}
The applications of gamma-ray are ubiquitous in our daily life, such as container security initiative\cite{lun2008institutional}, gamma-knife surgery\cite{ganz2012gamma}, nuclear medical imaging\cite{eisen1999cdte} and food storage\cite{LADO2002433}. While in the vastness of the universe, photons, ranging from several MeV to tens of TeV\cite{lamb1997point,abdo2010fermi,aharonian2004crab}, results from various different processes, such as energetic cosmic ray\cite{kulsrud1969effect,hunter1997egret}, luminous pulsars\cite{romani1996gamma} and gamma-ray burst\cite{gamma_ray_burst_RMP,gamma_ray_burst_theories}. The information of gamma-ray burst was firstly published by the results of Vela satellites\cite{klebesadel1973observations} and then were quickly verified by data from the Soviet satellites\cite{mazets1974cosmic}. The ability of cosmic sources to emit such intense gamma-rays indicates that investigating this extreme environment is a promising route to discover new physics which are impossible in earth-bound laboratories. 

An alternative method of generating violent emission of gamma-rays is through the interaction of petawatt ($10^{15}$W) lasers and plasmas in the laboratory. Several multi-PW laser facilities, such as Extreme Light Infrastructure (ELI)\cite{ELI} and Exawatt Center for Extreme Light Studies (XCELS)\cite{XCELS}, are expected to operate at intensities beyond 10$^{23}$W/cm$^{2}$ in next few years. Under  $\sim$10$^{23}$W/cm$^{2}$, various theoretical schemes have been put forward for multi-MeV photon sources with tens of percent for the total conversion efficiency, such as
reinjected electron synchrotron radiation\cite{Brady2012PRL}, skin-depth emission\cite{Ridgers2012PRL}, radiation reaction facilitating gamma-ray\cite{Nakamura2012PRL} and sandwich target design\cite{Stark2016PRL}.
Nevertheless, none of them has the ability to extend the energy of gamma photon up to several GeV, which is highly desirable to explore the laboratory astrophysics\cite{RevModPhys.78.755,RRD_bulanov2015}. Recently, exploiting the interplay between pair cascades\cite{bell2008possibility} and anomalous radiative trapping\cite{gonoskov2014anomalous}, ultrabright GeV photon source can be achieved in laser-dipole waves\cite{gonoskov2017ultrabright}. However, the scheme of dipole wave field\cite{gonoskov2012dipole,gonoskov2013probing,gonoskov2014anomalous} requires multi beams focused into a tiny point symmetrically, which is still an experimental challenge nowadays. Here we report an alternative all-optical scheme to realize the brilliant GeV gamma-ray emission via irradiating only one multi-PW circularly polarized (CP) pulse on a compound target in QED regime. This all optical backscatter scheme is already available in experiment for relative lower intensity circumstance\cite{phuoc2012all,chen2013mev,powers2014quasi,sarri2014ultrahigh,khrennikov2015tunable}.

\subsection{2. Theoretical model for coupling effect}
In the realm of nonlinear QED, electrons are able to emit a huge amount of kinetic energy in the form of high-energy photons $\gamma_{ph}$, as a result of absorbing a certain number $n$ of laser photons $\gamma_l$, $e^- + n\gamma_l \rightarrow e^- + \gamma_{ph}$. The invariant parameters $\eta=(e\hbar/m_e^3c^4)|F_{\mu\nu}p^\nu|=E_{RF}/E_{Sch}$ and $\chi=(e\hbar^2/2m_e^3c^4)|F_{\mu\nu}k^\nu|$ characterize the discrete photon emission process, where $e$ the electron charge, $m_e$ the electron rest mass, $\hbar$ the Planck constant, $c$ the light velocity in vacuum, $F_{\mu\nu}$ the field tensor and $p^\nu$ ($k^\nu$) the electron's (photon's) four-momentum. $E_{RF}$ denotes the electric field in the electron's rest frame and $E_{Sch}=m_e^2c^3/e\hbar\approx1.3\times10^{18}Vm^{-1}$ is the characteristic field of Schwinger limit\cite{schwinger1951gauge}. When $\eta\lesssim1$:(1) The radiation process should be described by probabilistic quantum emission rather than continuous one. (2) The corresponding quantum weaken correction for radiation is inevitable\cite{kirk2009pair,QED_domian_di2012}. When an electron beam co-propagates with the laser pulse, the electric force offset by the magnetic field effect results in $\eta\approx0$, which is undesired for high-energy photon emission\cite{RRD_bulanov2015,QED_domian_di2012}. However, if the laser counter-propagates with the electron beam, it leads to an enhancement as $\eta\approx 2\gamma E_L/E_{Sch}$, where $\gamma$ is the relativistic Lorentz factor of electron and $E_L$ is the polarized laser field. This colliding configuration can not only lower down the threshold of QED cascade from seed electrons\cite{grismayer2016laser}, but also facilitate the generation of $\gamma$-ray explosion\cite{nerush2011laser,gong2016high} and pair plasma\cite{bell2008possibility,zhu2016dense,jirka2016electron,chang2015generation}. 

\begin{figure*}[tbp]
\includegraphics[keepaspectratio=true,height=60mm]{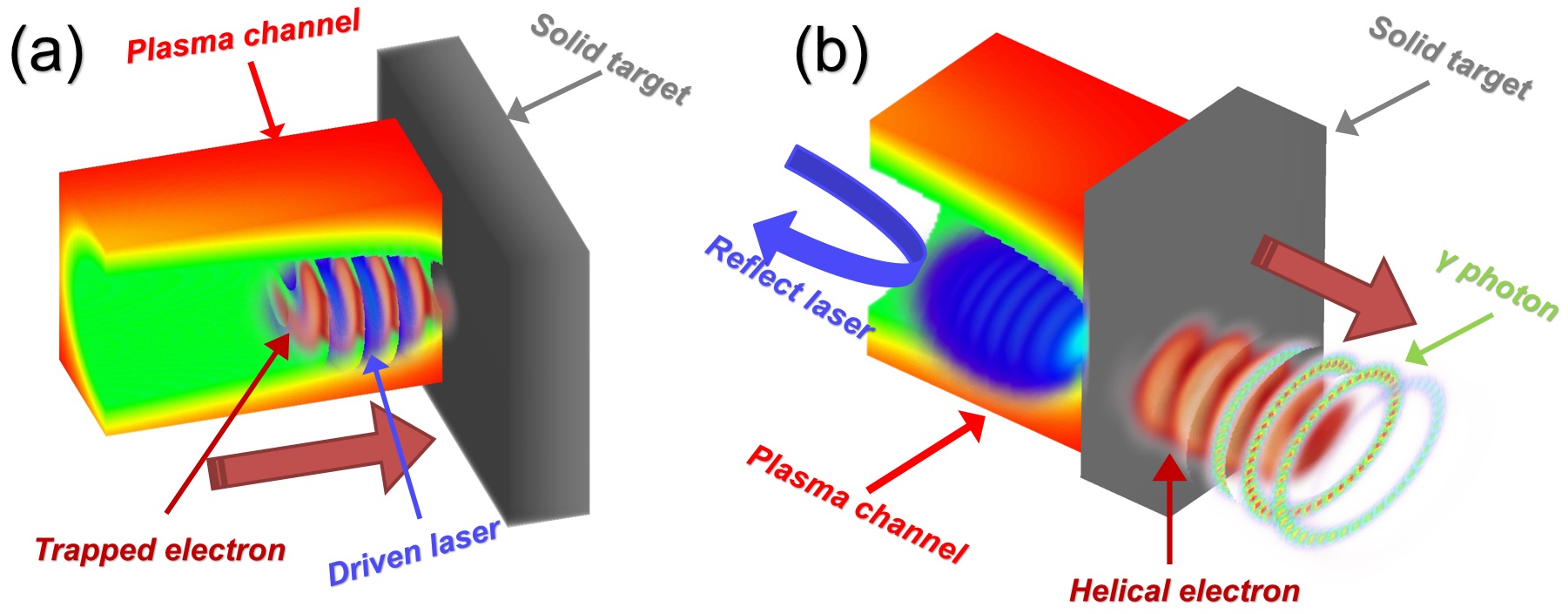}
\caption{Scheme of the ultra-brilliant GeV gamma-ray source with helical structure. (a) and (b) show light being reflected before and after respectively.}
\label{fig_schematic}
\end{figure*}

To exploit the counter-propagating configuration, in this letter, a CP femtosecond pulse was irradiated on a compound target (in Fig.\ref{fig_schematic}) consisted of a near-critical-density (NCD) plasma slab and a solid foil. Here the solid foil plays the role as a plasma mirror\cite{phuoc2012all,chen2013mev,powers2014quasi,sarri2014ultrahigh,khrennikov2015tunable} to spontaneously reflect the driven light to trigger the subsequent Compton backscattering. Generally, when a CP pulse of 10$^{19-21}$W/cm$^{2}$ propagates in the NCD target, the ionized electrons can be transversely expelled from central area to form a plasma channel\cite{pukhov1999channel,pukhov2002strong}. Some injected electron can experience a direct laser acceleration process and a collimated energetic electron bunch can be produced when its oscillation frequency in the channel field is close to the light frequency witnessed by the electron\cite{liu2013generating,hu2015dense,arefiev2012parametric}. However, under the higher intensity of $\sim$10$^{23}$W/cm$^{2}$, the injected electrons are mostly expelled from the central region and a hollow channel is merely filled with laser radiation\cite{ji2014radiation}. More interestingly, a great amount of electrons will be trapped back into the channel if radiation reaction (RR) is taken into account\cite{ji2014radiation,RR2016Chang}, where transverse ponderomotive force is properly balanced by the radiation recoil.

%the radial electrostatic field $\mathbf{E}_{Sr}=k_Er\hat{e}_r$, longitudinal electric field  $\mathbf{E}_{S\parallel}$ and quasistatic azimuthal magnetic field $\mathbf{B}_{S\theta}=-k_Br\hat{e}_\theta$ are generated in the excavated plasma channel\cite{pukhov1999channel,pukhov2002strong}. A collimated energetic electron beam with overcritical density and helical structure can be achieved in resonant regime\cite{liu2013generating,hu2015dense,arefiev2012parametric} when electron betatron frequency $\omega_{\beta}=\sqrt{e(v_\parallel k_B+k_E)/(\gamma m_e)}$ is close to the light frequency witnessed by the electron $\omega_L=(1-v_\parallel/v_{ph})\omega_0$. Here $v_{ph}$ is the phase velocity of light in plasma and $\omega_0$ is the light frequency in laboratory frame. However, under the higher intensity of $\sim$10$^{23}$W/cm$^{2}$, electrons are mostly expelled from the high-intensity region and a hollow channel is merely filled with laser radiation\cite{ji2014radiation}. More interestingly, a great amount of electrons will be trapped back into the channel if radiation reaction (RR) is taken into account\cite{ji2014radiation,RR2016Chang}, where transverse ponderomotive force is properly balanced by the radiation recoil.

It should be noted that the interaction between laser and NCD plasma is very complicated, where the filamentation instability\cite{honda2000collective}, hosing instability\cite{huang2017relativistic} or non optimal laser-plasma matching\cite{mourou2002design} can destroy the laser propagating and the channel's shape. Here a relatively large spot radius and the small plasma density are adopted to avoid these detrimental influence and guarantee the stable channel. To understand the underlying mechanism of RR impact on this scheme, the single electron model is utilized to depict the interaction with laser transverse field $E_L$ and self-generated fields in the plasma channel. Based on previous work\cite{pukhov2002strong,liu2015quasimonoenergetic,hu2015dense}, self-generated fields in the channel include radial electrostatic field $\mathbf{E}_{Sr}=k_Er\hat{e}_r$, longitudinal electric field  $\mathbf{E}_{S\parallel}$ and quasistatic azimuthal magnetic field $\mathbf{B}_{S\theta}=-k_Br\hat{e}_\theta$, where $k_E$ and $k_B$ can be seen as constant and are related to the plasma density. The time derivative of the ponderomotive phase $\psi$ can be written as

%the evolution of the relative phase $\psi$ between the electron rotation and the periodic laser field was explored the time derivative of the ponderomotive phase $\psi$ can be written as

\begin{equation}
\begin{aligned}\label{eq1}
\frac{d\psi}{dt}=\omega_{\beta}-\omega_L=\sqrt{\frac{e}{\gamma m_e}(v_\parallel \langle k_B\rangle+\langle k_E\rangle)}-(1-v_\parallel/v_{ph})\omega_0.
\end{aligned}
\end{equation}
Here $\omega_{\beta}=\sqrt{e(v_\parallel k_B+k_E)/(\gamma m_e)}$ is the electron betatron frequency and $v_\parallel$ ($v_\perp$) the electron longitudinal (transverse) velocity. $\omega_L=(1-v_\parallel/v_{ph})\omega_0$ is the Doppler-shifted laser frequency witnessed by electron, where $\omega_0$ is the laser frequency and $v_{ph}=c/\sqrt{1-\omega_p^2/(\gamma\omega_0^2)}$ is the laser phase velocity\cite{decker1995group}. $\omega_p$ is the plasma frequency. The $\psi$ is relative phase between the electron rotation and the periodic laser field. The time derivative of the electron Lorentz factor is expressed as
\begin{equation}
\begin{aligned}\label{eq2}
\frac{d\gamma}{dt}=\frac{-e\mathbf{E}\cdot\mathbf{v}-\mathbf{f_{rad}}\cdot\mathbf{v}}{m_ec^2}=-\frac{e( v_\perp E_L cos\psi+v_\parallel \langle E_\parallel\rangle)}{m_ec^2}-\epsilon_{rad}\omega_0\beta^2a_s^2\eta^2G(\eta).
\end{aligned}
\end{equation}
Here $E_L$ is the light electric field amplitude. Since the stochasticity of photon emission is difficult to be simplified into a precise formula, the discontinuous influence is neglected in the single model and the quantum corrected RR force $\textbf{f}_{rad}\approx-G(\eta)\epsilon_{rad}m_ec\omega_0\vec{\beta}a_{s}^2\eta^2$ is used in Eq.(\ref{eq2}) to qualitatively analyze the RR influences, where $G(\eta)\approx(1+12\eta+31\eta^2+3.7\eta^3)^{-4/9}$ is the quantum weaken factor\cite{kirk2009pair}. The impacts issued from the discrete stochasticity in RR is beyond the scope of this manuscript and these are worth discussing in the future work. $\epsilon_{rad}=4\pi r_e/3\lambda_0$ is the dimensionless ratio, where $r_e=e^2/m_ec^2\approx2.8\times10^{-15}m$ is the classical electron radius and $\lambda_0$ is the laser wavelength. $\vec{\beta}=\vec{v}/c$ is the normalized electron velocity and $a_{s}=eE_{Sch}/m_e\omega_0c$ is the normalized Schwinger field. The parameters in above equations depend on time and are probably in especially complicated form so that the average values denoted by $\langle\ \rangle$ are used. From Eqs.(1)-(2), it can be found that the phase space ($\psi$,$\gamma$) has a fixed point\cite{jordan2007nonlinear,hirsch2012differential} at ($\psi_0$,$\gamma_0$) = ($\cos^{-1}\frac{-\epsilon_{rad}\omega_0\beta^2a_s^2\eta^2G(\eta)m_ec^2-e v_\parallel \langle E_\parallel\rangle}{e v_\perp E_L},\frac{e(v_\parallel \langle k_B\rangle+\langle k_E\rangle)}{m_e(1-v_\parallel/v_{ph})^2\omega_0^2}$). To determine the system dynamic property from Eqs.(1)-(2) in ($\psi$,$\gamma$) space, the perturbation expansion nearby ($\psi_0$,$\gamma_0$) of Eqs.(1)-(2) was made and quadratic terms were dropped to approach the characteristic Jacobian matrix $\mathbf{Ja}$\cite{jordan2007nonlinear,hirsch2012differential}:
\begin{equation}
\begin{aligned}
\mathbf{Ja}\approx\begin{pmatrix} 0 & -\frac{1}{2}\sqrt{\frac{e}{\gamma^3m_e}(v_\parallel \langle k_B\rangle+\langle k_E\rangle)} \\ ev_\perp E_Lsin\psi & -\epsilon_{rad}m_ec^2\omega_0\beta^2a_{s}^2\frac{\partial G(\eta)\eta^2}{\partial\gamma} \end{pmatrix}_{\psi_0,\gamma_0}.
\end{aligned}
\label{eq3}
\end{equation}
Without RR effect, the trace and determinant of Jacobian matrix are tr(Ja)$=$0 and det(Ja)$>$0 when the right lower RR term is canceled, which manifests that ($\psi_0$,$\gamma_0$) is a center without any source or sink property\cite{jordan2007nonlinear,hirsch2012differential}. On the contrary, with RR effect included, at fixed point tr(Ja)$<$0 and det(Ja)$>$0 indicates that its behaviour converts from center to spiral sink attractor\cite{gonoskov2014anomalous,ESIRKEPOV20152044,gong2016radiation,kirk2016radiative}. The sink attractor emerging illustrates a large fraction of the radiation trapped electrons tends to possess the same relative phase $\psi_0$ with respect to laser electric field and the helical density structure is an intrinsic rotary manner of the electric field of CP laser. Due to electron moving in the same direction as the pulse, the electric field $E_L$ counteracts the force from laser magnetic field $B_L$ leading to $\eta\approx\gamma|\mathbf{E}_L+\mathbf{v}\times\mathbf{B}_L|/E_{Sch}\approx0$ and tr(Ja)$\sim$0. Notwithstanding, the strong self-generated magnetic field $B_{s\theta}\approx n_eR/(2\varepsilon_0c)$ (here $\varepsilon_0$ the permittivity of vacuum, $n_e$ the RR trapped electron density and $R$ the channel radius) approaching the order of driven laser field\cite{Stark2016PRL} gives $\eta\approx\gamma|\mathbf{E}_L+\mathbf{v}\times(\mathbf{B}_L+\mathbf{B}_{s\theta})|/E_{Sch}\approx\frac{\gamma B_{s\theta}}{E_{Sch}}$, which results in tr(Ja)$\approx-2\epsilon_{rad}\beta^2e^2B_{s\theta}^2/m_e\omega_0<$0 and enables the attractor effect on achieving such a helical electron bunch (HEB). The nearby electrons are attracted to possess the identical Lorentz factor $\gamma_0$=$e(v_\parallel \langle k_B\rangle+\langle k_E\rangle)/m_e(1-v_\parallel/v_{ph})^2\omega_0^2$. The total angular momentum (AM) along the longitudinal x-axis, i.e. $L=yp_z-zp_y$, acquired by the HEB can also be estimated as
\begin{equation}
{\centering \ L \approx -\int\Sigma_{i}er_\perp E_Lcos\psi_idt \ \ \ i=1,2,3...
}
\label{eq4}
\end{equation}
here r$_{\perp}$ is the electron transverse radius and the index i refers to the i-th electron. From Eq.(4) we can see that the laser could transfer its spin angular momentum (SAM) to HEB only when most of electrons possess the same ponderomotive phase $\psi_i$, otherwise ensemble average leads to $\sum_{i}cos\psi_i\approx$0. Therefore, coupling effects among RR trapping, self-generated magnetic field and spiral attractor in phase space, enhance the net AM gain and realize the HEB. Eventually the discrete photon emission\cite{ritus1985quantum,neitz2013stochasticity,blackburn2014quantum} is triggered through the inverse Compton scattering (ICS) between the HEB and reflected light, where prolific high-energy photons inheriting a large fraction of electrons' energy and AM are generated. 

\begin{figure}[tbp]
\includegraphics[keepaspectratio=true,height=80mm]{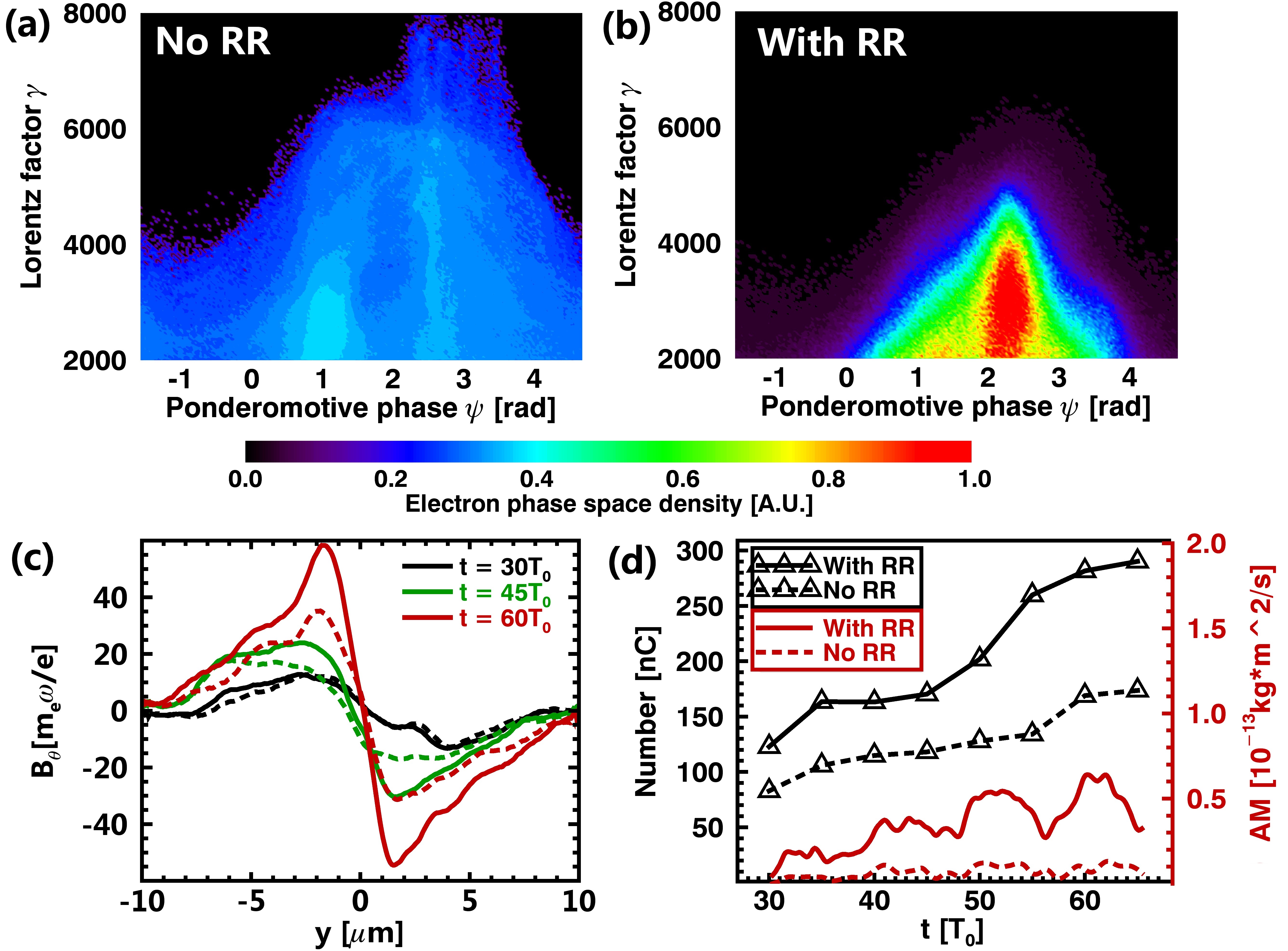}	\caption{Distribution of electron density in $\psi$-$\gamma$ space at t=50T$_0$ with RR (a) and without RR (b), respectively. (c) Normalized amplitude of $\mathbf{B}_{s\theta}$ averaged over the channel in the plane z=0 at time t=30,45,60 T$_0$, where solid (dash) line denotes the circumstance with (without) RR. (d)  presents the number of the electrons inside the channel and total AM of electrons in x direction as a function of the interaction time $t$, where solid (dash) corresponds to the case with (without) RR.}
\label{fig_attractor}
\end{figure}

\subsection{3. Particle-in-cell (PIC) simulation results}
The feasibility and robustness of this scheme are demonstrated by using the self-consistent three dimension PIC code EPOCH\cite{arber2015contemporary}.
A Monte Carlo probabilistic model\cite{duclous2010monte,ridgers2014modelling} has been successfully implemented, which is based on QED corrected synchrotron cross sections and coupled with the subsequent reduction of the electron momentum. Each particle is assigned an optical depth ($\tau$) at which it emits according to $P=1-e^{-\tau}$, where $P\in$[0,1] is chosen at random to consider the quantum correction in the emission processes as well as the straggling. The rates of photon production, $d\tau_\gamma/dt=(\sqrt{3}\alpha_fc\eta)/(\lambda_c\gamma)\int_{0}^{\eta/2}d\chi F(\eta,\chi)/\chi$, are then solved until the optical depth is reached, when the emission event occurs\cite{duclous2010monte}. Here, $\alpha_{fc}$ is the fine structure constant, $\lambda_c=\hbar/(m_ec)\approx3.9\times10^{-13}m$ is the Compton wavelength and $F(\eta,\chi)$ is the quantum synchrotron spectrum\cite{duclous2010monte}. % $\chi=(\hbar\omega_{ph}/2m_ec^2)|\mathbf{E}_\perp+\mathbf{\hat{k}}\times c\mathbf{B}|/E_{Sch}$, where the subscript $\perp$ denotes the component perpendicular to the unit vector in the photon's direction of motion $\mathbf{\hat{k}}$ and $\hbar\omega_{ph}$ is the photon energy.

The incident 1.2$\times$10$^{23}$W/cm$^{2}$ CP pulse propagates along X direction with a profile of $a$=$a_0e^{-(t-t_0)^4/\tau_0^4}e^{-(y^2+z^2)/r_0^2}sin(\omega_0t)$, where $\tau_0$=$5T_0$ denotes the intensity with a full width at half maximum (FWHM) of 25.6fs (T$_0$$\approx$3.3fs is the laser period) and $a_0$=$eE_L/m_e\omega_0c$$\approx$$300$ is the normalized amplitude of the laser field. $r_0$=$5\lambda_0$ is the spot size ($\lambda_0$=1.0$\mu m$). The simulation box is 80$\lambda_0 \times$ 40$\lambda_0 \times$ 40$\lambda_0$ in X $\times$ Y $\times$ Z direction, which has been uniformly divided into 3200 $\times$ 800 $\times$ 800 cells. A hydrogen slab with initial density of $n_e=2n_c$ locates between 10$\lambda_0$ to 60$\lambda_0$ and aluminum foil of $n_e=700n_c$ is placed from 60$\lambda_0$ to 80$\lambda_0$, where $n_c=m_e\omega_0^2/4\pi e^2$ is critical density\cite{gibbon2004short}. The hydrogen slab and aluminum foil contain 4 and 16 macroparticles per cell (for both species), respectively. For reference, there is no obvious difference in our results when we double the number of macroparticle per cell.

\begin{figure*}[tbp]
\includegraphics[keepaspectratio=true,height=100mm]{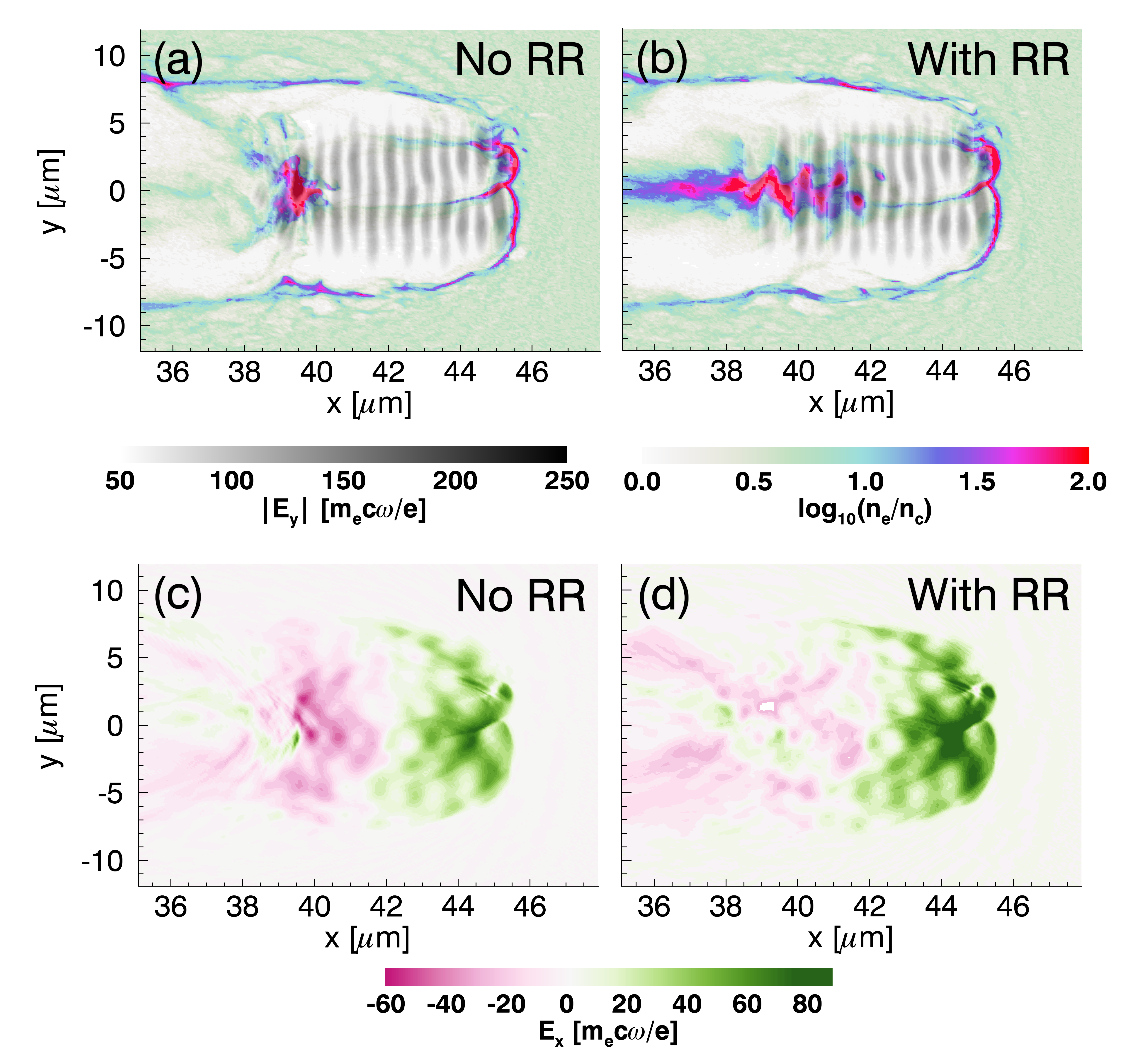}
\caption{(a) and (b) correspond to the distributions of electron density $n_e$ for the case without and with RR, where the absolute value of laser electric field $|E_y|$ is also figured in grey with a transparency of 60\%. The distributions of longitudinal field $E_x$ generated in the plasma channel are shown in (c)(d) as well.}
\label{fig_insert}
\end{figure*}

The electron density distributions in $\gamma-\psi$ space at t=50T$_0$ for the cases with and without RR are presented in Fig.\ref{fig_attractor}(a) and (b). Lorentz factor at the fixed point obtained from Eqs.(1)-(2) as $\gamma_0=\frac{e(v_\parallel\langle k_B\rangle+\langle k_E\rangle)}{m_e(1-v_{\parallel}/v_{ph})^2\omega_0^2}\approx\frac{(v_\parallel/c)(n_e/n_c)}{2[1-v_\parallel/c\sqrt{1-n_e/(a_0n_c)}]^2}$ where $\langle k_B\rangle\approx\frac{en_e}{2\epsilon_0}$,$\omega_0=\sqrt{\frac{n_ce^2}{\epsilon_0m_e}}$ and $v_{ph}\simeq\frac{c}{\sqrt{1-n_e/(a_0n_c)}}$ are taken into account and $\langle k_E\rangle$ is neglected as the transverse static electric field is relatively weak compared with self-generated magnetic field. Substituting $n_e=2n_c$, $a_0=300$ and $v_\parallel=0.9863c$ (from simulation parameters and results) into above equation leads to $\gamma_0=3416$. Considering $\epsilon_{rad}=1.18\times10^{-8}(\frac{1\mu m}{\lambda_0})$, $\beta\approx1$, $a_s\approx4.1\times10^5$, $\eta\approx\frac{\gamma_0B_{s\theta}}{a_s}\approx0.165$, $G(\eta)\approx1$, $\langle E_\parallel\rangle\approx0.015E_L$ and $v_\perp=0.165c$, the relative phase is deduced as $\psi_0=\arccos\frac{-\epsilon_{rad}\omega_0\beta^2a_s^2\eta^2G(\eta)m_ec^2-ev_\parallel\langle E_\parallel\rangle}{ev_\perp E_L}\approx2.24$. When RR force is switched on, most of electrons possess a relative phase $\psi$=2.3 in Fig.\ref{fig_attractor}(b) which is in good agreement with our theoretically derived attractor point ($\psi_0, \gamma_0$)=(2.24, 3416). Since neither RR trapping nor attractor emerging occurs, the number density of electron in Fig.\ref{fig_attractor}(a) is relatively small compared to RR case and it does not behave like the attractor modulated distribution. The self-generated azimuthal magnetic field B$_{s\theta}$ averaged over the channel cross plane z=0 is plotted in Fig.\ref{fig_attractor}(c) with maximum $\approx$0.6MT (normalized value equals 60$m_e\omega_0/e\approx$0.2B$_L$, where B$_L$ is the laser magnetic amplitude) at t=65T$_0$, which demonstrates that RR recoil enhances the $B_{s\theta}$ generation due to the more trapped electron current along longitudinal axis. This kind of self-generated magnetic field in channel can not only enhance the gamma photon emission\cite{Stark2016PRL}, but also help accelerate ions in the rear surface of target\cite{bulanov2015helium}, which has already been verified in experiment under lower laser intensity with shock-compressed gas target\cite{helle2016laser}. The temporal evolution of electron number inside the plasma channel and their total AM $L=\sum_i y_ip_{zi}-z_ip_{yi}$ are recorded in Fig.\ref{fig_attractor}(d) for both cases. It is found that RR not only boosts the electron accumulation inside the channel but also facilitates the AM transfer to HEB, which is in good agreement with the theoretical prediction of Eq.(4). The RR force prevents electrons from being expelled transversely, resulting in a increase of electrons from 172 nano-Coulombs(nC) to 291 nC at t=65T$_0$. The enhancement of electron current strengthens the B$_{s\theta}$, which gives a positive feedback on spiral attractor merging in phase space and effectively favors angular momentum transformation from laser's SAM to HEB's AM.

The electron density distributions for the case with and without RR are shown in Fig.3(a)(b). Here the emergence of helical spatial structure depends on the RR impact, which accords with the attractor facilitating electron density modulation with the similar frequency as laser electric field in Eq.(\ref{eq3}). When RR is switched off, a ball of electrons are injected into the tail of plasma channel and can be accelerated by the longitudinal electric field $E_x$. The distributions of $E_x$ are plotted in Fig.\ref{fig_insert}(c)(d) for with RR case or not. Since the quantity of electron in the channel for RR case is much higher than that for no RR, the sheild effect weakens the accelerating field in RR case when compared to the no RR one.

\begin{figure*}[tbp]
\includegraphics[keepaspectratio=true,height=65mm]{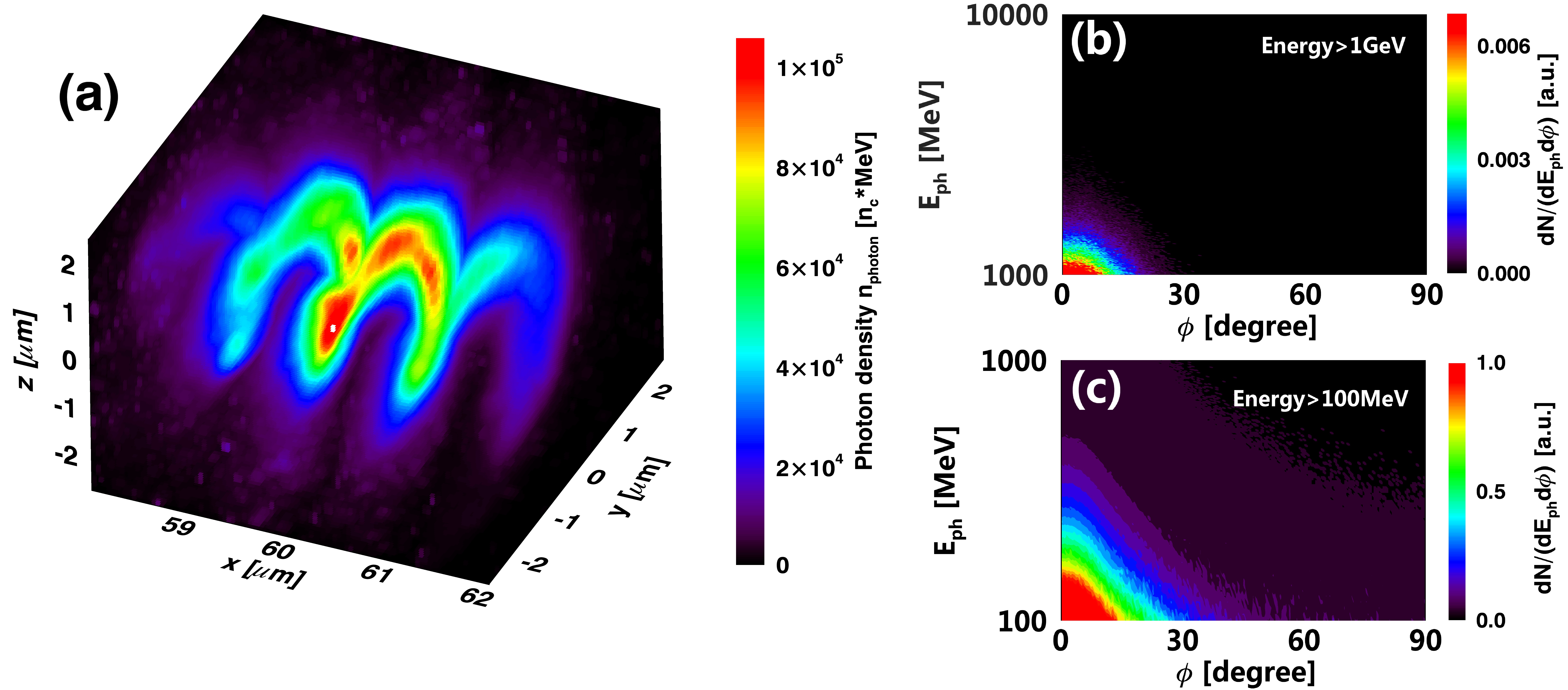}
\caption{(a) Volume distribution of the photon energy density where only photon with energy higher than 10MeV is recorded for less computation costs. (b)(c) Final photon angular-spectral distribution for energy higher than 1GeV and 100MeV respectively.}
\label{fig_photon}
\end{figure*}

\begin{figure*}[tbp]
	\includegraphics[keepaspectratio=true,height=60mm]{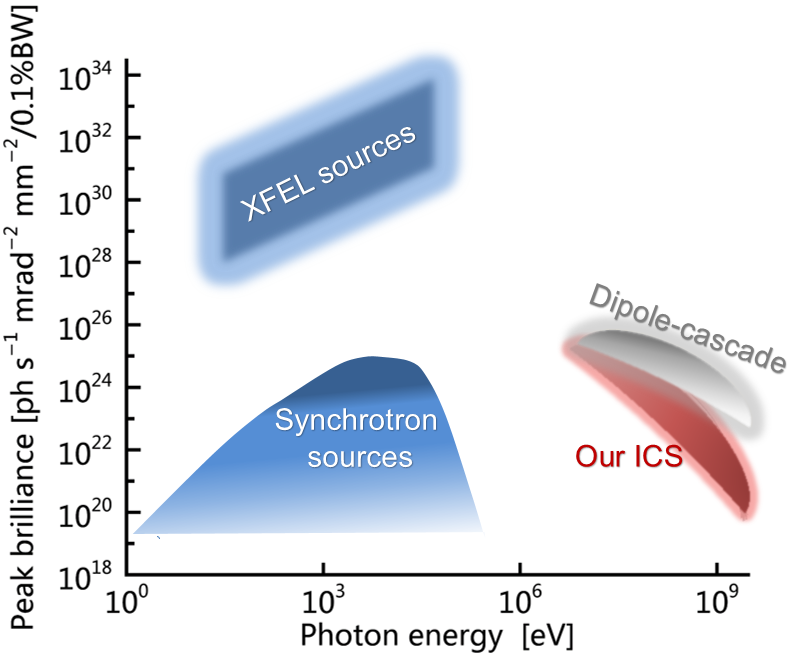}
	\caption{Comparison of the peak brilliance of our proposed ICS source with the other existing photon source, e.g., Synchrotron, XFEL and Dipole-cascade\cite{gonoskov2017ultrabright}.}
	\label{fig_brill}
\end{figure*}

Since the ponderomotive force of CP pulse avoids the longitudinal oscillation at twice the optical frequency\cite{gibbon2004short}, plasma in the second layer cannot be heated violently and the driven light is substantially reflected. Under colliding configuration, the parameter $\eta\approx2\gamma E_L/E_{Sch}\gtrsim1$ indicates that the discrete incoherent photon emission\cite{QED_domian_di2012} gives a more appropriate description compared with the coherent electromagnetic wave radiation derived from the Li\'{e}nard-Wiechert retarded potential\cite{jackson1999classical}. The volume snapshot of the photon energy density at t=70T$_0$ is exhibited in Fig.\ref{fig_photon}(a) where photon beam inherits spatial helical structure and transverse size of the source is about 1.5$\mu$m. The gamma-ray flash duration is $\sim$16fs roughly equal to half of the laser because the driven pulse and trapped electrons completely overlap inside the channel. The angular-spectral distribution calculated by accumulating the forward photons at t=70T$_0$ over the entire simulation region is shown in Figs.\ref{fig_photon}(b) and (c). Most of energetic photons are highly collimated and predominantly located within an emission polar angle $\phi\leq$15$^\circ$ ($\phi\leq$30$^\circ$) for energies higher than 1GeV (100MeV). In a 0.1\% bandwidth (BW) around 1GeV we have 1.05$\times$10$^8$ photons, implying the brightness of 1.7$\times$10$^{23}$ photons/s/mm$^2$/mrad$^2$/0.1\%BW for the GeV gamma-ray emission. The corresponding source brilliances at 100 MeV and 10 MeV are 2.3$\times$10$^{24}$ and 1.5$\times$10$^{25}$ photons/s/mm$^2$/mrad$^2$/0.1\%BW, respectively. The comparasion among different photon source is illustrated in Fig.\ref{fig_brill}. Our ICS scheme predominantly aims at high brilliance around GeV. Another dipole wave field can achieve the brightest gamma photon emission with 9$\times$10$^{24}$ photons/s/mm$^2$/mrad$^2$/0.1\%BW at GeV\cite{gonoskov2017ultrabright}, but the dipole wave needs to be realized through symmetrically colliding multi pulses, which is still a challenge in experiment. Here our scheme shooting one laser pulse onto double layer target is the most efficient method to generate brilliant GeV gamma ray source\cite{Brady2012PRL,Ridgers2012PRL,Nakamura2012PRL,Stark2016PRL} and it is more experimentally accessible.

\subsection{4. Discussion and conclusion}
\begin{figure}[tbp]
	\includegraphics[keepaspectratio=true,height=80mm]{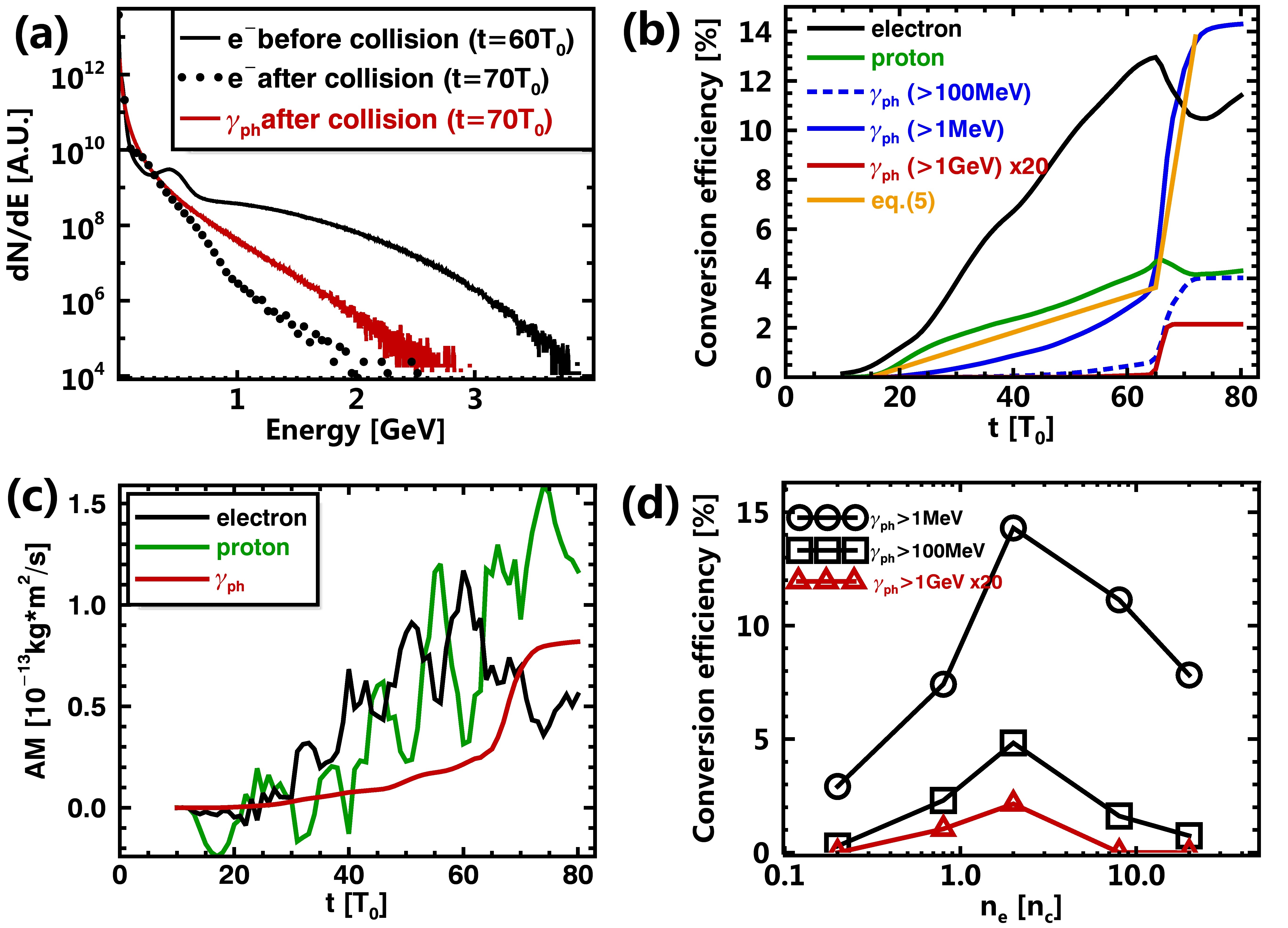}
	\caption{(a) The energy spectra of electrons at t=60,70T$_0$ and photons at t=70T$_0$. (b) The laser energy conversion to the electrons (black), protons (green) and gamma-ray photons ($>$1MeV in solid blue, $>$100MeV in dash blue and $>$1GeV in solid red). The photon with energy greater than 1GeV, rendering in red, corresponds to the right red axis. The orange solid line plots the theoretical radiation prediction from eq.(4). (c) Temporal evolution of the total AM of electrons, protons and photons ($>$1MeV). (d) The laser energy conversion to $\gamma$-photons with different plasma densities. Here the value of $\gamma_{ph}>$1GeV is times by 20 and horizontal (density) axis is on logarithmic scale.}
	\label{fig_discussion}
\end{figure}

In Fig.\ref{fig_discussion}(a), the exponential decay spectrum of photon covers higher energy range from 1MeV to several GeV with a cutoff energy at 2.9GeV and that of the electron before(t=60T$_0$) and after(t=70T$_0$) ICS process are presented. The nonlinear QED regime predicts that most photons are emitted with an energy $h\nu_{ph}\approx$0.44$\eta\gamma m_ec^2$\cite{bell2008possibility,kirk2009pair} which carries a large fraction of electron's kinetic energy. It is obvious that the amount of high-energy electron is drastically curtailed with the cutoff-energy declining from 3.9GeV to 2.5GeV and simultaneously most of energy is converted to gamma photons. The temporal evolutions of the particle energy are illustrated in Fig.\ref{fig_discussion}(b), where 14.5\%;4.2\%;0.108\% of the total laser energy is transferred into the gamma-ray photon with energy above 1MeV;100MeV;1GeV. For energies above 100MeV and 1GeV, the photons are emitted almost exclusively by ICS process during 65T$_0<$t$<$70T$_0$. Based on power radiated by a single electron, $P_{rad}=(4\pi m_ec^3/3\lambda_c)\alpha_fc\eta^2G(\eta)$\cite{duclous2010monte,ridgers2014modelling}, the instantaneous radiation power of this regime can be estimated as
\begin{equation}
	{\centering P_{rad}\approx\left\{
		\begin{aligned}
			N_e\frac{4\pi\alpha_fm_ec^3}{3\lambda_c}(\frac{\gamma B_{s\theta}}{E_{Sch}})^2G(\frac{\gamma B_{s\theta}}{E_{Sch}})   &\ \ \ & t<t_{ref}, \\
			N_e\frac{4\pi\alpha_fm_ec^3}{3\lambda_c}(\frac{2\gamma E_L}{E_{Sch}})^2G(\frac{2\gamma E_L}{E_{Sch}})   &\ \ \ & t\geq t_{ref}.
		\end{aligned}
		\right.
	}
	\label{eq5}
\end{equation}
Here t$_{ref}$=65T$_0$ is the time of light reflecting and $\eta$ is approximated by $\gamma B_{s\theta}/E_{Sch}$ at $t<t_{ref}$ and $2\gamma E_L/E_{Sch}$ at $t\geq t_{ref}$, respectively. The length of NCD plasma $l=50\mu m$ is not comparable with the laser depletion length $L_{depletion}\approx c\tau_0a_0n_c/n_e$=750$\mu m$\cite{pukhov2002strong,lu2007generating}, as a result a large part of laser energy is reflected and backscatter with the electron bunch. In addition, when laser propagates in the NCD plasma, both of its intensity and spot size will change due to the self-focusing, self-modulation, etc. The radius of the self-generated channel is defined by the balance of the ponderomotive and charge separation fields. Here, we choose the laser spot almost the same as the radius of such channel. That results in no significant change of the laser transverse size during the propagation in near critical plasma and we assume they are constant in estimation of Eq.(5). Eq.(5) predicts the radiation power P$_{rad}\approx$0.63PW (t$<$t$_{ref}$) and P$_{rad}\approx$19.2PW (t$\geq$t$_{ref}$) which qualitatively agrees with the simulation results in Fig.\ref{fig_discussion}(b), implying the nonlinear QED ICS based gamma-ray source power of the same order as the infrared incident laser.

The transfer of axial AM from the laser to the particles is plotted in Fig.\ref{fig_discussion}(c). The oscillation of electron and proton AM is due to charged particles interplaying with the laser electromagnetic field. The different sign of electric charge causes the opposite oscillation direction in electron and proton. Since the spiral attractor results in the fixed relative phase between electron velocity and laser electric field, the overall AM of electron rises gradually before backscattering with the reflected pulse. However, photons do not interplay with laser field and their AM has a moderate growth before the ICS. The photons are predominantly emitted from electron modulated by the spiral attractor during 65T$_0<$t$<$70T$_0$ so that a sharp photon AM increase and a pronounced electron AM drop occur in ICS process. In terms of quantum mechanics, the angular momentum carried by a photon of CP laser is $\sigma=\pm1$ for spin motion. The total angular momentum absorbed from laser is approximately expressed as $L_{l}=\delta\frac{W_{l}}{\hbar\omega_0}\sigma\hbar$=1.70$\times$10$^{-12}\delta$ kg$\ast$m$^2$/s, where $W_{l}$ is the whole laser energy and $\delta$ is the absorbing ratio. During ICS process, AM is more efficiently transferred from electron to gamma-ray and eventually the AM of photons reaches 8.2$\times$10$^{-14}$ kg$\ast$m$^2$/s, 4.8\% of the total laser SAM. In addition, a parameter scan has been carried out to investigate energy conversion efficiency for a wide range density 0.2-20n$_c$ of first layer plasma with thickness of 50$\mu$m in Fig.\ref{fig_discussion}(d) and find that there is an optimal condition $n_e\sim n_c$ for realizing the twisted GeV gamma-ray emission. The disadvantage for relatively rarefied plasma (n$_e$=0.2n$_c$) is lack in trapped helical electron amount so that insufficient electron quantity accounts for deficient gamma-ray production, while for relatively dense circumstance (n$_e\gtrsim$10n$_c$) driven laser tends to deplete most of their energy in the first slab and without any remnants to trigger ICS process.

In conclusion, we have shown how the ultra-intense and ultra-bright GeV gamma-ray flash can be achieved by irradiating a prospective 1.2$\times$10$^{23}$W/cm$^{2}$ laser on a compound plasma target in nonlinear QED regime. The initial energetic HEB results from the coupling effects among RR trapping, self-generated magnetic field and emergency of spiral attractor in $\gamma$-$\psi$ space. The helical gamma-ray flash inherits a considerable AM and energy of the parent electron beam through Compton backscattering between HEB and the reflected driven pulse. The final photon source, with unprecedented power of 20 PW and brightness of 1.7$\times$10$^{23}$ photons/s/mm$^2$/mrad$^2$/0.1\% (at 1 GeV), might enable significant development of application in particles physics and laboratory astrophysics. Our scheme is also feasible in the laboratory system where cluster jets\cite{fukuda2009energy} or nano-tube foams\cite{ma2007directly} can be utilized for NCD plasma generation and a solid foil acts as a plasma mirror to reflect laser. Such parameters of the gamma-ray sources will be achieved with the next generation of multi-PW laser facilities in the future.

\section*{Acknowledgements}
The work has been supported by the National Basic Research Program of China (Grant No.2013CBA01502), NSFC (Grant Nos.11535001) and National Grand Instrument Project (2012YQ030142). The PIC code Epoch was in part funded by the UK EPSRC grants EP/G054950/1, EP/G056803/1, EP/G055165/1 and EP/M022463/1. J.Q. Yu wants to thank the Project (2016M600007,2017T100009) funded by China Postdoctoral Science Foundation. Our simulations were carried out in Max Planck Computing and Data Facility and Shanghai Super Computation Center. The author Z.Gong acknowledges fruitful discussions with Prof. S.V. Bulanov and H.X. Chang.

\bibliography{aa}

\end{document}